\documentclass[citeautoscript,prx,aps,twocolumn,showpacs,showkeys,superscriptaddress,amsmath,amssymb]{revtex4}

\usepackage{graphicx}
\usepackage{psfrag}
\usepackage{epsfig}
\usepackage{color}
\usepackage{subfigure}
\definecolor{dred}{rgb}{0.7,0.0,0.0}
\usepackage{dcolumn}
\usepackage{bm}

\begin{document}

%
%
\title{Possible Bicollinear Nematic State with \\
Monoclinic Lattice Distortions in Iron Telluride Compounds}

\author{Christopher B. Bishop}
\author{Jacek Herbrych} 
\author{Elbio Dagotto}
\author{Adriana Moreo}
\affiliation{Department of Physics and Astronomy, University of Tennessee,
Knoxville, TN 37966, USA} 
\affiliation{Materials Science and Technology Division,
Oak Ridge National Laboratory, Oak Ridge, TN 37831, USA}

\date{\today}

\begin{abstract}
{Iron telluride (FeTe) is known to display bicollinear magnetic order at low temperatures
together with a monoclinic lattice distortion. Because the bicollinear order can involve two
different wavevectors $(\pi/2,\pi/2)$ and $(\pi/2,-\pi/2)$, symmetry considerations allow for the
possible stabilization  of a nematic state with short-range bicollinear order 
coupled to monoclinic lattice distortions at a $T_S$ higher than 
the temperature  $T_N$ where long-range bicollinear order fully develops. As a concrete example,
the three-orbitals spin-fermion model for iron telluride is studied with an additional 
coupling  $\tilde\lambda_{12}$ between the monoclinic lattice strain
and an orbital-nematic order parameter with $B_{2g}$ symmetry.
Monte Carlo simulations show that with increasing  $\tilde\lambda_{12}$ the first-order
transition characteristic of FeTe splits and   
bicollinear nematicity is stabilized 
in a (narrow) temperature range. In this new regime the lattice is
monoclinically distorted and short-range spin and orbital 
order breaks rotational invariance. A
 discussion of possible realizations of this exotic state is provided.}

\end{abstract}
 

 
\maketitle

\section{Introduction} 

The theoretical understanding of high critical temperature superconductivity in iron compounds has evolved
from its early qualitative developments based on Fermi surface nesting to more quantitative efforts incorporating
the role of electronic correlations~\cite{intro,HKM,kotliar,dainat,RMP2013,FCS}. 
In particular, experts have focused on several
complex  regimes including 
electronic nematicity~\cite{fisher2010,fisherother,fernandes1,fernandes2}, 
an interesting state observed in several high critical temperature pnictide 
superconductors~\cite{clarina,rotundu,kasahara,zhou}.
Upon cooling, this nematic phase is reached at a temperature $T_{S}$, 
concomitantly with a structural phase transition from a tetragonal
to an orthorhombic lattice. Upon further cooling a magnetically ordered phase is stabilized at a lower 
temperature $T_{N}$. The orthorhombic nematic 
phase between $T_{S}$ and $T_{N}$ exhibits a reduced symmetry 
under rotations from $C_4$ to $C_2$. This is also observed in the magnetic 
and orbital degrees of freedom leading to nonzero magnetic and orbital ``nematic''
order parameters. Experimental investigations have shown that 
this nematic phase occurs in the parent compounds of the 1111 pnictides~\cite{clarina}.
Since the orthorhombic lattice distortion $\delta_O=|a_O-b_O|/(a_O+b_O) \sim 0.004$~\cite{huang} is 
small ($a_O$ and $b_O$ are the lattice parameters in the orthorhombic notation), it is often 
argued that the lattice 
plays the role of a ``passenger'' in the nematic transition which is believed to be driven by either 
the magnetic or orbital degrees of freedom. In addition, it is interesting to notice that the structural 
transition occurs simultaneously with the N\'eel temperature in 
several other iron-based materials. For example, members of the 122 family need to be electron 
doped, with the chemical replacement occurring
directly on the FeAs planes, to develop the nematic phase~\cite{rotundu,kasahara,zhou}. 
Hole doping, or electron doping via chemical substitution away from the FeAs planes, fails to establish nematicity~\cite{chenx,saha}.

\begin{figure}[thbp]
\begin{center}
\includegraphics[width=8cm,clip,angle=0]{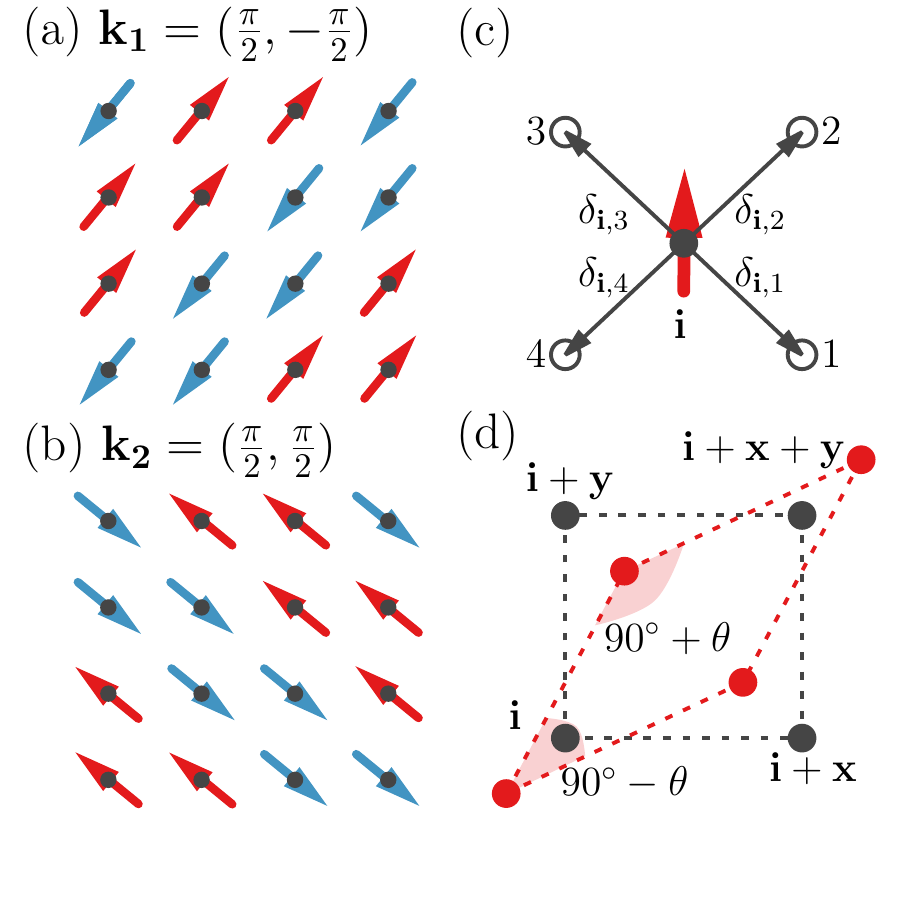}
\vskip -0.3cm
\caption{(color online) (a) The bicollinear antiferromagnetic spin order with 
wavevector $(\pi/2,-\pi/2)$. 
(b) same as (a) but for the state lattice-rotated 
by 90 degrees with wavevector $(\pi/2,\pi/2)$. 
(c) Schematic drawing of an iron atom at site ${\bf i}$ 
(filled symbol) and its four Te neighbors (open symbols),
projected in the $x$-$y$ plane 
in their equilibrium position. The distances $\delta_{{\bf i},\nu}$ between the irons at site ${\bf i}$ 
and its four neighboring Te atoms are
indicated as well. The localized spin ${\bf S}_{\bf i}$ is also sketched.
(d) Schematic drawing of the Fe lattice
equilibrium position in the tetragonal phase (black symbols and lines) and in the monoclinic phase (red symbols and lines). 
Four Fe atoms are indicated
with filled symbols and labeled by their lattice site index.} 
\vskip -0.4cm
\label{diagnem}
\end{center}
\end{figure}

In the chalcogenides, the parent compound FeTe exhibits an unexpected ``bicollinear'' 
magnetic state~\cite{bao,li,CMR}, shown in panels (a,b) 
of Fig.~\ref{diagnem}, whose $T_{N}$ coincides with the $T_{S}$ of a structural transition 
to a phase with a monoclinic lattice distortion, as shown 
in panel (d) of the same figure. This joint transition is strongly first order~\cite{bao,chen,fobes}. 
The reported lattice distortions in Fe$_{1.076}$Te
and Fe$_{1.068}$Te are $\delta_M=|a_M-b_M|/(a_M+b_M) \sim 0.007$ ~\cite{bao} ($a_M$ and $b_M$ are the low-temperature 
lattice parameters in the monoclinic notation). Replacing Te with Se the bicollinear magnetic order is eventually lost, 
the material becomes superconducting, and it develops an orthorhombic 
nematic phase above its superconducting critical temperature. 
In recent theoretical work, using a spin-fermion model we explained 
the bicollinear magnetic order using symmetry considerations 
as a consequence of the monoclinic distortion~\cite{bistripes,bnl2016}. 
Based on this reasoning, the role of the lattice in the case of FeTe
appears more important than previously anticipated.

The aim of the present work is to argue that the pnictides and chalcogenides 
could potentially behave more symmetrically with
regards to the presence of a nematic state. As expressed above, the pnictides 
either already have nematicity without doping, as in
the 1111 compounds, or develop nematicity 
after doping as in the Co-doped 122 compounds. Based on symmetry arguments, the presence
of a nematic regime is theoretically understood as follows.
In these materials the magnetic ground state has wavevector $(\pi,0)$, with staggered spins
along the $x$-axis and parallel spins along the $y$-axis. However, the $(0,\pi)$ state 
should have the same energy by symmetry. In cases of two-fold degeneracy in the ground state, it was
predicted that an Ising transition could occur upon cooling~\cite{premi}, 
with an order parameter that breaks lattice
rotational invariance and involves only short-range magnetic correlations. 
Upon further cooling, the O(3) full symmetry breaking process is possible. 

Our main observation here is that the bicollinear state shown in Fig.~\ref{diagnem}~(a) with wavevector 
${\bf k_1} = (\pi/2,-\pi/2)$ has a partner, displayed in Fig.~\ref{diagnem}~(b), with identical 
energy but ${\bf k_2} = (\pi/2,\pi/2)$~\cite{peaks}.
Then, the same Ising-O(3) rationale expressed above for the $(\pi,0)-(0,\pi)$ degeneracy can be repeated 
for bicollinear states: starting at high temperature both spin structure factors 
$S({\bf k})$  will start growing 
with equal strength upon cooling at the wavevectors ${\bf k_1}$ and ${\bf k_2}$. 
By analogy with the pnictides, 
it is possible that at a critical nematic
temperature $T_{S}$ an asymmetry develops such that $S({\bf k_1}) > S({\bf k_2})$, 
and then at a lower temperature
$T_{N}$, $S({\bf k_2})$ drops to zero while $S({\bf k_1})$ grows like the volume.

While no nematic phase with these characteristics has been reported yet in
materials of the FeTe family with the bicollinear spin order, the 
present study provides computational evidence that there are Hamiltonians 
with spin- and orbital-lattice coupling that display this  new nematic behavior if the 
couplings strengths are properly tuned. While our many-body tools do not allow us to 
predict what specific material may display this 
phenomenon, our symmetry arguments and concrete
simulation results are offered as motivation 
for the experimental search for this exotic bicollinear-nematic state.

Previous numerical studies of spin-fermion models for pnictides 
with spin, orbital, and lattice degrees of freedom provided indications that the structural transition
is due to the coupling between the lattice and spins~\cite{shuhua13}. 
Thus, in these regards 
the lattice follows the spins. But the spin-lattice coupling leads to 
$T_S = T_N$ and, then, the establishment of a nematic phase with $T_S > T_N$ requires a 
more subtle mechanism. Investigations by our group have shown that the  nematic regime can be achieved by 
the addition of an orbital-lattice coupling~\cite{shuhua13} (or by
the introduction of in-plane magnetic disorder, namely by replacing iron 
by non-magnetic atoms~\cite{chris15,fernandes-dis}). Based on this previous research, here 
a coupling between the monoclinic lattice distortion
and an {\it orbital} nematic parameter with $B_{2g}$ symmetry will be added 
to the spin-fermion model that already has the spin-lattice coupling previously 
developed to study FeTe~\cite{bistripes}. It will be shown below that this addition generates
the novel bicollinear nematic state. This is not an obvious result because the tight-binding term contains
an intrinsic tendency towards collinear magnetic order that could have affected the fragile nematicity
region that we are reporting here. 

The publication is organized as follows. In Section~\ref{model} the model is described including the
new term that must be incorporated in order to stabilize a bicollinear-nematic state.
In Section~\ref{methods} we provide an explanation of the numerical approach that allows for the
parallelization of the Monte Carlo procedure and the concomitant use of clusters 
of reasonable size for our purposes. The main results showing the stabilization of the new nematic state
are presented in Section~\ref{results}. The discussion, including 
possible physical realizations, is in Section~\ref{disc}, with brief conclusions in Section~\ref{conclu}.

\section {Models}\label{model}
 
In the first subsection we will discuss a general model that addresses the interactions between electrons and the
lattice for the case of monoclinic distortions. In the second subsection, the actual special case that was 
computationally investigated in this publication will be presented. Our effort aims to prove that there is at least
one set of couplings for which, varying temperature, a bicollinear nematic state is stabilized. A more comprehensive analysis
of phase diagrams varying the many couplings in the generic model would demand considerable computational resources
and this task is left for future investigations. 

\subsection{Generic Model}

The most general spin-fermion (SF) Hamiltonian discussed here is an extension of the purely 
electronic model previously introduced~\cite{BNL,shuhua}, 
supplemented by additional couplings to the monoclinic lattice degrees of freedom~\cite{shuhua13,chris,igor}:
\begin{equation}
H_{\rm SF} = H_{\rm Hopp} + H_{\rm Hund} + H_{\rm Heis} + H_{\rm Stiff}+ H_{\rm SLM}+ H_{\rm OLM}.
\label{ham}
\end{equation}
\noindent $H_{\rm Hopp}$ represents the three-orbitals ($d_{xz}$, $d_{yz}$, $d_{xy}$) 
tight-binding Fe-Fe hopping of electrons, with the
hopping amplitudes selected to reproduce photoemission data (see Eqs.(1-3) and Table 1 
of~\cite{three}). In the undoped-limit the average electronic density per iron and per orbital 
is set to $n$=4/3~\cite{three} and a chemical potential in $H_{\rm Hopp}$~\cite{chris} controls its value. 
The on-site Hund interaction is $H_{\rm Hund}$=$-{J_{\rm H}}\sum_{{\bf i},\alpha} {{{\bf S}_{\bf i}}\cdot{{\bf s}_{{\bf i},\alpha}}}$,
where ${{\bf S}_{\bf i}}$ are the localized spins at site ${\bf i}$  
and ${\bf s}_{{\bf i},\alpha}$ are spins corresponding to orbital $\alpha$ of the itinerant fermions at the same site.
For computational simplicity, the localized spins are assumed classical and of norm one~\cite{foot}.
$H_{\rm Heis}$ contains the nearest neighbor (NN) and next-NN (NNN) Heisenberg interactions among the localized spins, 
with respective couplings $J_{\rm NN}$ and $J_{\rm NNN}$. 
As explained before~\cite{shuhua13,shuhua}, both NN and NNN could be active 
because of the geometry of the problem, where in each layer the Te atoms (or As, Se, P) 
are at the centers of iron plaquettes as seen from above.
However, in our previous study of FeTe~\cite{bistripes} we observed
that the experimental value of $T_N$ for FeTe could be obtained 
by simply setting $J_{\rm NN} = J_{\rm NNN} = 0$. This is due to the fact that the 
intersite spin-spin couplings favor either 
checkerboard ($J_{\rm NN}$) or collinear ($J_{\rm NNN}$) 
magnetic configurations and in order to obtain
a bicollinear ground state it is necessary to use 
a larger value of the spin-lattice coupling 
$\tilde g_{12}$ which, in turn, increases $T_N$~\cite{intersite}.
$H_{\rm Stiff}$ is the lattice stiffness given by a Lennard-Jones potential 
to speed up convergence~\cite{chris} (full expression can be found in~\cite{shuhua13}). 

Recently, an important novel term was introduced~\cite{bistripes} to describe FeTe properly.
This term has the form $H_{\rm SLM}$=$-g_{12}\sum_{\bf i}\Psi_{NNN}({\bf i})\epsilon_{12}({\bf i})$ and it 
provides a coupling between the localized spins and the monoclinic $\mathcal{M}_{\rm ono}$ lattice 
distortions~\cite{kuo}. 
The coupling constant strength is $g_{12}$ and the spin NNN nematic order parameter is defined as
\begin{equation}
\Psi_{NNN}({\bf i}) ={1\over{2}} { {{\bf S}_{\bf i}}\cdot{
({\bf S}_{\bf i + \bf x + \bf y}+{\bf S}_{\bf i - \bf x - \bf y}-
 {\bf S}_{\bf i+\bf x - \bf y}-{\bf S}_{\bf i-\bf x+\bf y}) }},
\label{PsiNNNdef}
\end{equation}
\noindent where ${\bf i}\pm\mu\pm\nu$ indicates the four NNN sites to ${\bf i}$, with $\mu=\pm{\bf x}$  
and $\nu=\pm{\bf y}$ representing unit vectors along the $x$ and $y$ axes, respectively. 
Note that $\Psi_{NNN}({\bf i})$ has the
value 2~(-2) in the perfect bicollinear states shown in Figs.~\ref{diagnem}~(a) and (b), respectively 
characterized by a peak at wavevectors $(\pi/2,-\pi/2)$ and $(\pi/2,\pi/2)$
in the magnetic structure 
factor. $\epsilon_{12}({\bf i})$ is the lattice $\mathcal{M}_{\rm ono}$ strain 
defined in terms of the Fe-Te distances $\delta_{{\bf i},\nu}$ as
\begin{equation}
\epsilon_{12}({\bf i})={1\over{8}}(|\delta_{{\bf i},2}|+|\delta_{{\bf i},4}|-|\delta_{{\bf i},1}|-|\delta_{{\bf i},3}|),
\label{e12}
\end{equation}
\noindent where $\delta_{\bf i,\nu}=(\delta^x_{\bf i,\nu},\delta^y_{\bf i,\nu})$
($\nu$=1,...,4) is the distance between Fe
at site ${\bf i}$ and each of its four Te neighbors 
(see panel (c) of Fig.~\ref{diagnem} and also Fig.~S1, Suppl. Sec. of ~\cite{bistripes}).   
As in previous simulations, the Te atoms are allowed to move locally from their equilibrium position only along
the $x$ and $y$ directions because the $z$ component of the position does not couple to the monoclinic distortion. 
It is important to notice that both $\Psi_{NNN}({\bf i})$ and 
$\epsilon_{12}({\bf i})$ transform according to the $B_{2g}$ representation of the $D_{\rm 4h}$ symmetry 
group, which means that
the spin-lattice term of the Hamiltonian transforms as $A_{1g}$ as expected. 
As the spin-lattice coupling $g_{12}$ grows and induces
a monoclinic $\mathcal{M}_{\rm ono}$ 
distortion, $\Psi_{NNN}$ develops a nonzero expectation value leading to the bicollinear spin
state order as explained in~\cite{bistripes}.

The Hamiltonian described thus far~\cite{bistripes} leads to a first-order phase
transition where both the monoclinic lattice and the bicollinear spin orders develop simultaneously. Thus, no bicollinear-nematic
state was reported in~\cite{bistripes}. Based on previous investigations of pnictides 
using the spin-fermion model~\cite{shuhua13}, it is natural
to introduce a coupling between the lattice and the {\it orbital} degree of freedom to favor nematicity (note
that adding this term does not imply immediately that the bicollinear nematic state will be stabilized because there are
competing collinear tendencies in the tight-binding term; a specific calculation is thus needed, as presented below). 
For the new term, care with regards to the symmetry of the operators used is required.
The monoclinic orbital-nematic order parameter is defined as
\begin{equation}
\Phi_{B_{2g}}({\bf i})=n_{{\bf i},XZ}-n_{{\bf i},YZ}=\sum_{\sigma} 
(c^{\dagger}_{{\bf i},xz,\sigma} c^{\phantom{\dagger}}_{{\bf i},yz,\sigma}-
 c^{\dagger}_{{\bf i},yz,\sigma} c^{\phantom{\dagger}}_{{\bf i},xz,\sigma}),
\label{PhiB2gdef}
\end{equation}
\noindent where $n_{{\bf i},\beta}$=$\sum_{\sigma}c^{\dagger}_{{\bf i},\beta,\sigma}
c^{\phantom{\dagger}}_{{\bf i},\beta,\sigma}$ ($\beta = XZ,YZ$), 
and the $B_{2g}$ orbital 
basis is related to the $B_{1g}$ orbital basis by
\begin{equation}
c_{{\bf i},XZ,\sigma}={1\over{\sqrt{2}}}(c_{{\bf i},xz,\sigma}+c_{{\bf i},yz,\sigma})
\label{cXZ}
\end{equation}
\noindent and
\begin{equation}
c_{{\bf i},YZ,\sigma}={1\over{\sqrt{2}}}(c_{{\bf i},xz,\sigma}-c_{{\bf i},yz,\sigma}).
\label{cYZ}
\end{equation}
Notice that the $x$ and $y$ axes point along nearest-neighbor irons, i.e. along the sides of the plaquette formed 
by four irons, while the $X,Y$ axes point along next nearest-neighbor iron, i.e.
along the diagonals of the iron plaquette. The $Z$ and $z$ axis coincide and they are perpendicular 
to the plane formed by the iron layer. 

The new term in the Hamiltonian $H_{\rm OLM}$ that couples the $B_{2g}$ orbital and lattice order parameters is given by
\begin{equation}
H_{\rm OLM}=-\lambda_{12}\sum_{\bf i}\Phi_{B_{2g}}({\bf i})\epsilon_{12}({\bf i}).
\label{Holm}
\end{equation}
Because the monoclinic lattice distortion $\epsilon_{12}({\bf i})$ transforms as the $B_{2g}$ representation of $D_{\rm 4h}$, it
must be coupled to an orbital order parameter that also transforms as $B_{2g}$ which is why 
$\Phi_{B_{2g}}({\bf i})$ was constructed. This ensures that
$H_{\rm OLM}$ is invariant under the $D_{\rm 4h}$ symmetry group. 

\subsection{Parameter Space Studied}

Although the model described thus far is generic for the spin-fermion family of Hamiltonians,
including Heisenberg couplings as well as lattice-spin and lattice-orbital terms,
in practice we have setup to zero some of those couplings for simplicity. The reason is
that we aim to prove computationally the existence of the novel proposed bicollinear nematic state
at least in the most optimal region of parameter space. In practice, it would be impossible 
to establish the full phase diagram varying every single parameter in the complete model, but as
experiments searching for the novel phase progress we can refine our analysis in future efforts.

$H_{\rm SF}$ was studied here with the same Monte Carlo (MC) procedure
employed in~\cite{shuhua13}, supplemented with the recently developed ``Parallel Travelling Cluster 
Approximation (PTCA)''~\cite{PTCA} described in the next section. The particular 
values for the couplings $J_{\rm H}=0.1$ eV, $J_{\rm NN}=J_{\rm NNN}=0$, 
and $\tilde g_{12}={2g_{12}\over{\sqrt{kW}}}=0.24$  
were chosen because they provide $T_{N}=T_{S}=70$~K 
for $\lambda_{12}=0$~\cite{bistripes}, which is the transition 
temperature experimentally observed in FeTe (note then that the Heisenberg couplings are neglected in
this first exploratory study, for simplicity). The coupling strength $\tilde g_{12}$ is the dimensionless version 
of the spin-lattice coupling, employing $W=3$~eV as the bandwidth of the tight-binding term 
and $k$ as the constant that appears in $H_{\rm Stiff}$~\cite{shuhua13}. Since these couplings were discussed
extensively before, in the present effort we will instead
focus on a careful description of the new dimensionless monoclinic orbital-lattice coupling 
$\tilde\lambda_{12}={2\lambda_{12}\over{\sqrt{kW}}}$ and its effects.

During the simulation the Te atoms are allowed to move locally away from their equilibrium 
positions within the $x$-$y$ plane. The Fe atoms can move globally
via a monoclinic distortion $\mathcal{M}_{\rm ono}$ where the angle between two orthogonal Fe-Fe bonds 
is allowed to change globally to $90^o+\theta$ with the four angles
in the iron plaquette adding to $360^o$, so that the next 
angle in the plaquette becomes $90^o-\theta$, with $\theta$ as
a small angle (see Fig.~\ref{diagnem}~(d)). In addition,
the localized spins ${\bf S_i}$ 
and atomic displacements $(\delta^x_{{\bf i},\nu},\delta^y_{{\bf i},\nu})$ 
that determine the value of the local $\mathcal{M}_{\rm ono}$ lattice 
distortion $\epsilon_{12}({\bf i})$~\cite{bistripes} (see Fig.~\ref{diagnem}~(c))
are evaluated via a standard Monte Carlo procedure.

\section {Methods: the Parallel Traveling Cluster Approximation }\label{methods}

\begin{figure}[thbp]
\begin{center}
\includegraphics[width=6cm,clip,angle=-90]{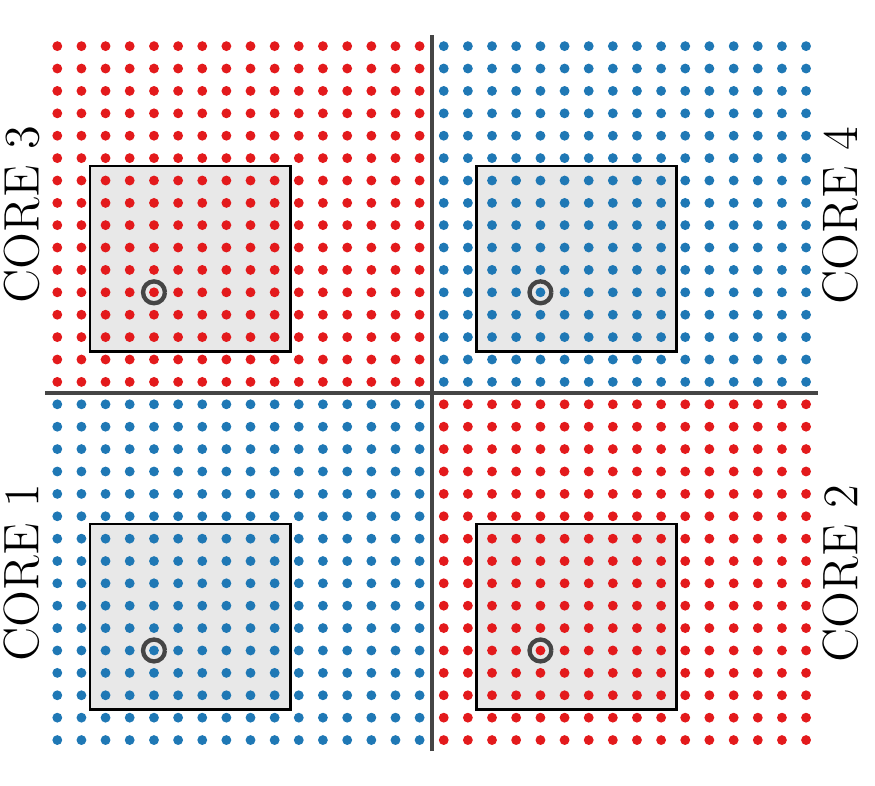}
\vskip -0.3cm
\caption{(color online) Diagram of the PTCA set-up used to sample the local spin and lattice variables. 
The lattice is divided into four quadrants and each of four processors 
generates traveling clusters (indicated with 8$\times$8 
squares) and proposes updates for the 
sites (indicated by small open circles) inside one quadrant.}
\vskip -0.4cm
\label{PTCA1}
\end{center}
\end{figure}

To access the lattice sizes needed to study the existence of a monoclinic nematic phase we 
implemented the Parallel Traveling Cluster Approximation (PTCA)~\cite{PTCA} which is a parallelization 
improvement over the traveling cluster approximation (TCA) previously introduced~\cite{tca}. 
PTCA allows parallelization in order 
to use multiple CPU cores and by this procedure we can reach lattices as large as $32 \times 32$. 
To perform a Monte Carlo update of one of the local variables -- either the localized spin ${\bf S_i}$ at the iron 
site ${\bf i}$ or the local distortion of the Fe-Te bonds joining the Fe atom at site ${\bf i}$ 
with its four Te neighbors --
an $8\times 8$ traveling cluster is constructed around site ${\bf i}$ and the Hamiltonian is diagonalized only 
inside that cluster to decide whether the update is accepted. The algorithm is parallelized by dividing 
the lattice into four quadrants with $16\times 16$ sites, one per different CPU core. 
Then, each CPU generates traveling $8\times 8$ clusters around the sites belonging to its quadrant, 
see Fig.~\ref{PTCA1} for an illustration, and these clusters are then simultaneously diagonalized.  

\begin{figure}[thbp]
\begin{center}
\includegraphics[width=6cm,clip,angle=-90]{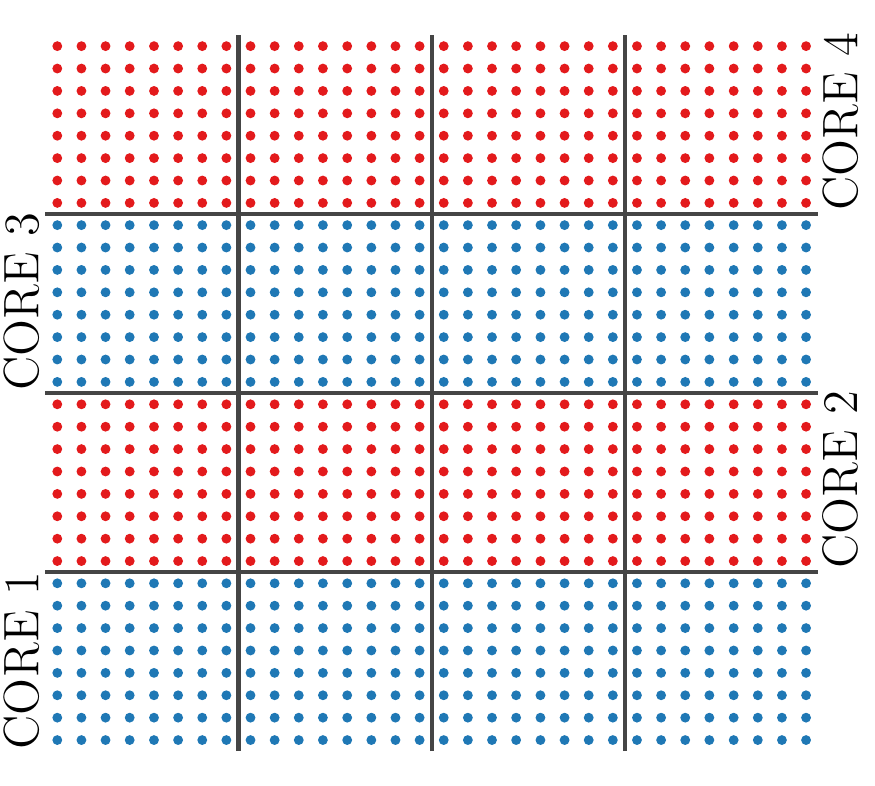}
\vskip -0.3cm
\caption{(color online) Diagram of the PTCA set-up used to sample the global lattice distortion variables. The lattice is divided into sixteen clusters. 
Each of the four processors diagonalizes four of the clusters.}
\vskip -0.4cm
\label{PTCA2}
\end{center}
\end{figure}

To update the global monoclinic lattice distortion given by the angles in the rhombus formed by the 
four irons shown in Fig.~\ref{diagnem}~(d) an extra new modification in the PTCA was introduced.
The $32 \times 32$ sites lattice was divided into 16 clusters with $8\times 8$ sites each as shown in 
Fig.~\ref{PTCA2}. Each of four CPU cores was devoted to diagonalize four of the clusters 
as indicated in the figure. The same update is proposed in all the clusters 
which are simultaneously diagonalized. Then, all the eigenvalues are collected in one of 
the cores in order to calculate the probability of the Monte Carlo update and decide
whether the update is accepted or rejected. 

For thermalization typically 5,000 
Monte Carlo steps were used, while 10,000 to 25,000 steps were performed in between 
measurements for each set of parameters and temperatures. The spin-spin correlation functions in real space 
were measured and the magnetic structure factor $S(k_x,k_y)$ was calculated via their Fourier transform.
Notice that in the bicollinear state the magnetic structure factor diverges for $(k_x,k_y)=(\pi/2,\pi/2)$ 
or $(\pi/2,-\pi/2)$. The N\'eel temperature $T_{N}$ 
is obtained from the magnetic susceptibility given by
\begin{equation}
\chi_{S(k_x,k_y)}=N\beta\langle S(k_x,k_y)-\langle S(k_x,k_y)\rangle\rangle^2,
\label{msus1}
\end{equation}
\noindent where $\beta=1/k_BT$ and $N$ is the number of lattice sites. We also calculated the numerical derivative of $S(\pi/2,\pi/2)$ with respect to 
temperature to double-check the value of $T_{N}$. 
The monoclinic structural transition temperature, $T_{S}$, 
was obtained by calculating the structural susceptibility given by   
\begin{equation}
\chi_{\delta_M}=N\beta\langle\delta_M-\langle\delta_M\rangle\rangle^2,
\label{msus2}
\end{equation}
\noindent where $\delta_M\approx\theta/2$ and $\theta$ is the deviation from 90$^o$ of the angle of the lattice plaquette as shown in  
Fig.~\ref{diagnem}~(d)~\cite{bistripes}. 
$T_{S}$ was also obtained from the numerical derivative of $\delta_M$ as a function of 
temperature and from monitoring
the behavior of the spin-nematic and orbital-nematic 
order parameters, $\Psi_{NNN}({\bf i})$ and $\Phi_{B_{2g}}({\bf i})$ respectively, 
introduced in the previous section and their associated susceptibilities.

\section {Results}\label{results}

As explained before, in previous work~\cite{bistripes} we found that the magneto-structural 
transition experimentally observed in FeTe, with $T_{S} =T_{N}=70$~K, was reproduced by setting $J_{\rm H}=0.1$~eV 
and $\tilde g_{12}=0.24$, and by dropping the Heisenberg couplings i.e. using $J_{\rm NN}=J_{\rm NNN}=0$.
In the present study for simplicity we keep fixed the values of all these parameters while 
we only vary the orbital-lattice coupling $\tilde\lambda_{12}$ 
to investigate whether a nematic phase can be stabilized in a range of temperature. Future work
will address what occurs in other portions of parameter space, such as with finite Heisenberg couplings
[some partial results are already available (to be shown in future publications) and all indicates that the bicollinear nematic
state is still present at large $\tilde\lambda_{12}$ even including nonzero $J_{\rm NN}$ and $J_{\rm NNN}$].

\subsection {Special case $\tilde\lambda_{12}=1$}

Similarly as with the behavior reported before for the spin-fermion model 
in the case of the pnictides with $(\pi,0)$ spin order~\cite{shuhua13},
in the bicollinear case studied here it was also observed 
that the novel bicollinear nematic region becomes stable
by increasing the value of the orbital-lattice coupling. This is not obvious because
of competing collinear tendencies, as already explained.
Another similarity with the case of the collinear state~\cite{shuhua13} 
is that the addition of the orbital-lattice
coupling $\tilde\lambda_{12}$ turns the first order magnetic transition into a second order one.
The temperature width of nematicity remains narrow, as in many previous investigations, 
and robust values of $\tilde\lambda_{12}$
are required. Nevertheless, this is sufficient to demonstrate the matter-of-principle existence of the bicollinear-nematic state
discussed in this publication.
For clarity, first let us address in detail the largest value of the coupling that we 
studied which was $\tilde\lambda_{12}=1$.

\begin{figure}[thbp]
\begin{center}
\includegraphics[width=9.5cm,clip,angle=0]{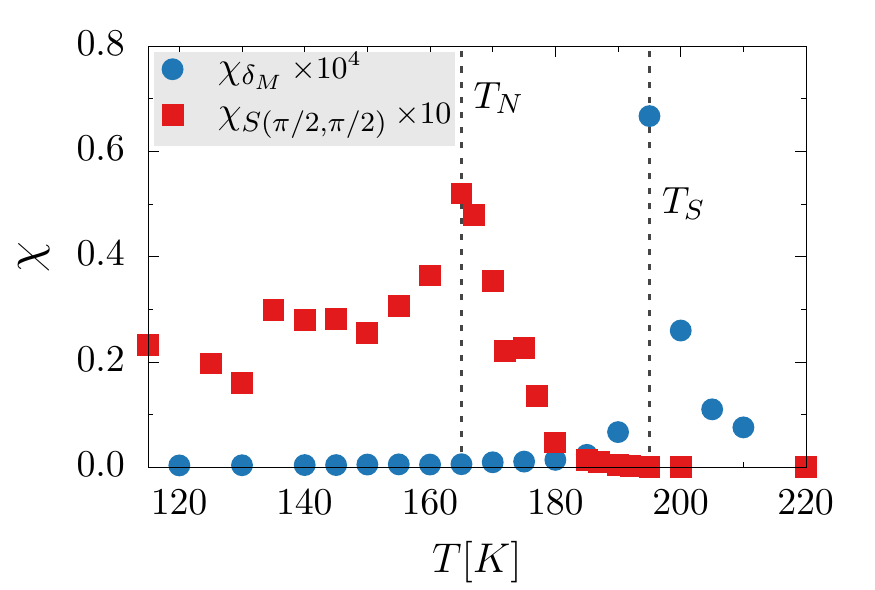}
\vskip -0.3cm
\caption{(color online) Magnetic susceptibility $\chi_S$ (squares) and monoclinic lattice susceptibility
$\chi_{\delta_M}$ (circles) evaluated using the PTCA algorithm at $\tilde\lambda_{12}=1$ 
employing a $32\times 32$ sites cluster. In this plot, 
and other plots of susceptibilities shown below, the fluctuations between subsequent
temperatures are more indicative of the error bars than the intrinsic errors bars
of individual points, which for this reason are not shown.}
\vskip -0.4cm
\label{chi1}
\end{center}
\end{figure}

In Fig.~\ref{chi1} the magnetic susceptibility $\chi_{S(\pi/2,\pi/2)}$ versus temperature is shown. A clear maximum at 
$T_{N}=165$~K indicates the magnetic transition to the bicollinear state with long-range order. 
The monoclinic lattice susceptibility is also shown. Interestingly, this quantity has a sharp 
peak at a clearly larger temperature $T_{S}=193$~K where the structural transition from 
tetragonal to monoclinic takes place,
indicating that a bicollinear-nematic state does indeed occur. 

\begin{figure}[thbp]
\begin{center}
\includegraphics[width=9.5cm,clip,angle=0]{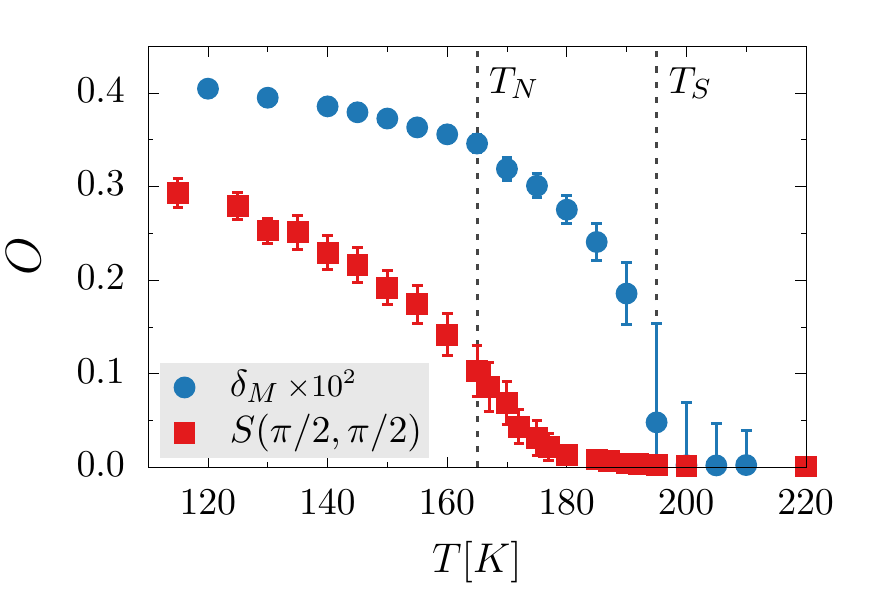}
\vskip -0.3cm
\caption{(color online) Magnetic spin structure factor $S(\pi/2,\pi/2)$ (squares) 
and monoclinic lattice order parameter $\delta_M$
(circles) evaluated using the PTCA algorithm for $\tilde\lambda_{12}=1$ on a $32\times 32$ sites cluster.}
\vskip -0.4cm
\label{psi1}
\end{center}
\end{figure}

In Fig.~\ref{psi1} the magnetic structure factor at wavevector 
$(\pi/2,\pi/2)$ is displayed. The $T_{N}$ from the susceptibility, 
shown with a dashed line, should occur 
when the rate of increase of the order parameter is maximized. 
This has been verified by performing a spline fit of the $S(\pi/2,\pi/2)$ 
points obtained 
from the Monte Carlo simulation and taking the numerical derivative.
The monoclinic lattice order parameter $\delta_M$ 
is also presented in Fig.~\ref{psi1}. The 
structural transition temperature is displayed with a dashed line as well. We also 
verified that the maximum in the lattice susceptibility from Fig.~\ref{chi1} coincides
with the maximum rate of change in the lattice order 
parameter via a spline fit of the Monte Carlo data.

\begin{figure}[thbp]
\begin{center}
\includegraphics[width=8.5cm,clip,angle=0]{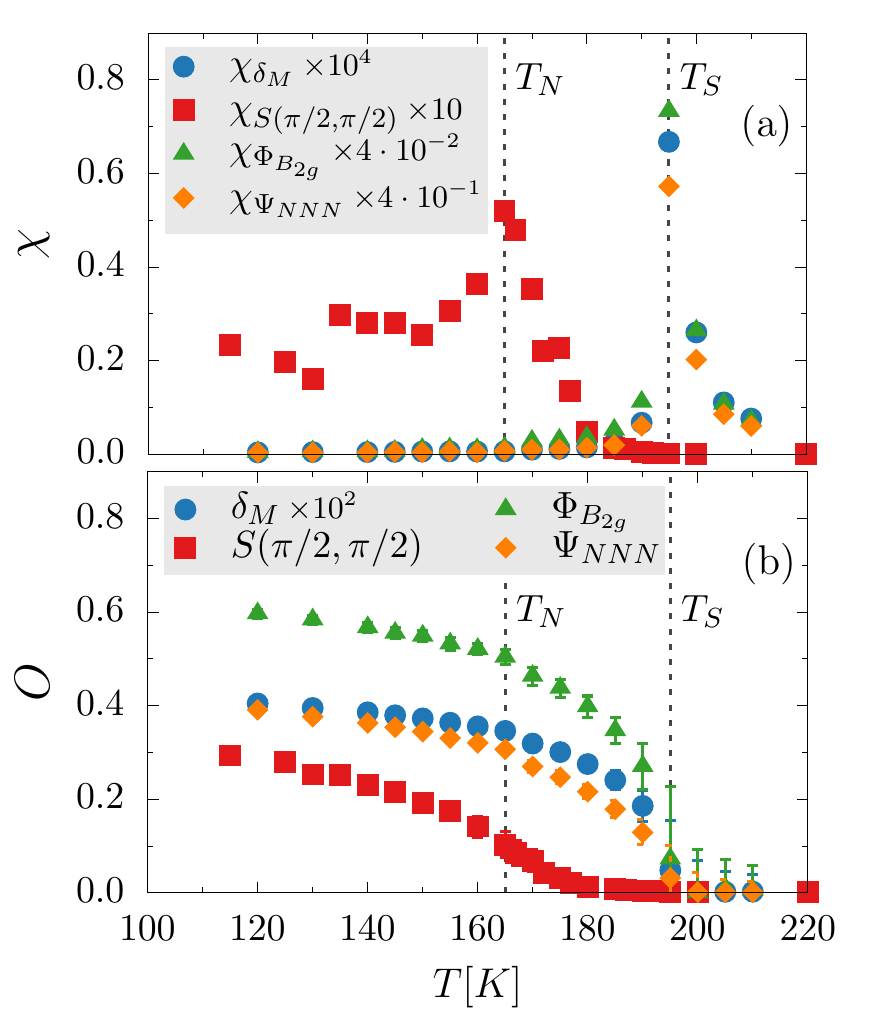}
\vskip -0.3cm
\caption{(color online) (a) Magnetic susceptibility $\chi_{S(\pi/2,\pi/2)}$ (red squares) 
with a maximum at $T_{N}=165~$K (dashed line), and the monoclinic lattice 
susceptibility $\chi_{\delta_M}$ (blue circles), 
spin-nematic susceptibility $\chi_{\Psi}$ (orange diamonds), and orbital-nematic susceptibility $\chi_{\Phi}$ 
(green triangles) all with a maximum at $T_{S}=193~$K. The susceptibilities were calculated 
at $\tilde\lambda_{12}=1$ using $32\times 32$ lattices. (b) Monte Carlo 
measured order parameters associated to (a). Shown are the magnetic structure 
factor $S(\pi/2,\pi/2)$ (red squares), monoclinic lattice distortion $\delta_M$ (blue circles),
spin-nematic  order parameter $\Psi_{NNN}$ (orange diamonds), and
orbital-nematic order parameter $\Phi_{B_{2g}}$ (green triangles).
The transition temperatures were obtained from
the susceptibilities in (a)  and via numerical derivatives in (b). Both procedures give the same result.}
\vskip -0.4cm
\label{orbnem}
\end{center}
\end{figure}

In between the two transition temperatures $T_{N}$ and $T_{S}$, a nematic phase 
is stabilized. In this phase both short-range orbital and spin 
nematic order develop
as it can be seen in Fig.~\ref{orbnem}, where 
in panel (a) the susceptibilities associated with various order parameters are presented. It can be observed 
that the orbital-nematic and spin-nematic susceptibilities 
have maxima at $T_{S}$ as does the structural susceptibility. This confirms 
the presence of a monoclinic nematic phase characterized by orbital-nematic and spin-nematic orders. 
These properties are also reflected in the behavior of the respective order parameters shown in panel (b) of the figure.

Performing spline fits of the order parameters and taking numerical derivatives, the critical temperatures 
obtained from the susceptibilities were reproduced.
It is important to notice that the lattice distortions $\delta_M \sim 10^{-3}$ are quantitatively similar to those reported 
in FeTe experiments while, as shown in Fig.~\ref{orbnem}~(b), the 
orbital and spin nematic order parameters develop values an order of magnitude larger. 
Thus, the strength of the orbital-lattice coupling used still leads to small lattice distortions but appears to generate
robust magnetic and orbital short-range order inducing substantial anisotropic effects in these observables.

\subsection {Special case $\tilde\lambda_{12}=0.85$}

\begin{figure}[thbp]
\begin{center}
\includegraphics[width=8.5cm,clip,angle=0]{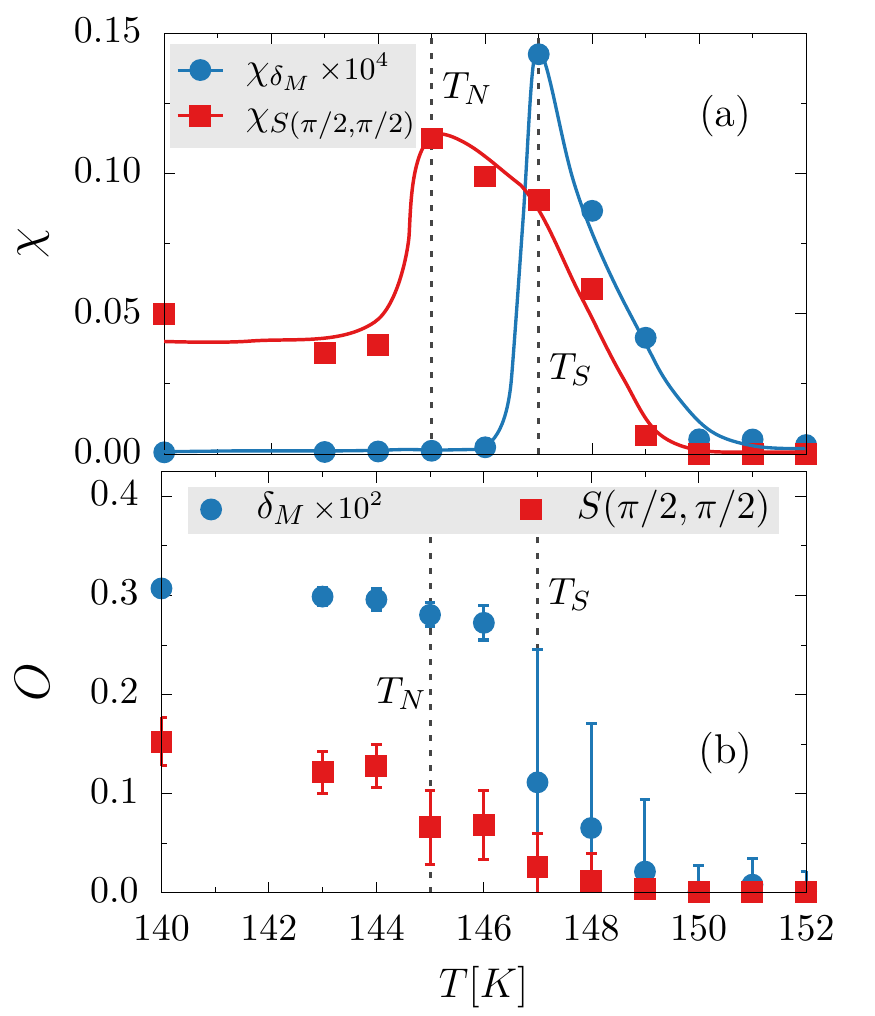}
\vskip -0.3cm
\caption{(color online) (a) Susceptibilities associated with the 
magnetic spin structure factor $S(\pi/2,\pi/2)$ (squares) and with the monoclinic lattice distortion (circles)
using $\tilde\lambda_{12}=0.85$ and a $32\times 32$ cluster. Solid lines are guides to the eye.
(b) Spin structure factor  $S(\pi/2,\pi/2)$ (squares) and monoclinic lattice order parameter $\delta_M$
(circles) for the same $\tilde\lambda_{12}$ and cluster size as in (a).}
\vskip -0.4cm
\label{lam85}
\end{center}
\end{figure}

As the value of the orbital-lattice coupling is reduced 
the separation between the magnetic and the structural transitions decreases. 
In panel (a) of Fig.~\ref{lam85} the magnetic and structural susceptibilities 
at $\tilde\lambda_{12}=0.85$ obtained from Monte Carlo simulations are presented. 
In this case $T_{N}=145$~K while $T_{S}=147$~K. 
The orbital- and spin-nematic order parameters also have a maximum 
susceptibility at $T_{S}$ (not shown for simplicity).
The magnetic and structural order parameters are
shown in panel (b) of Fig.~\ref{lam85} and their qualitative behavior is in
agreement with panel (a).
The indicated transition temperatures have been obtained from numerical fits of
the order parameters and their
derivatives as described in the previous subsection. This case $\tilde\lambda_{12}=0.85$
is close to the limit of our numerical accuracy. In principle, 
it is possible that simulations using larger systems and with far more
statistics may unveil a very narrow bicollinear nematic state even for 
small values of $\tilde\lambda_{12}$. However, 
for our qualitative purposes simply showing the stability of the new 
proposed phase in any range of $\tilde\lambda_{12}$ is sufficient.

\subsection {Phase Diagram}
 
\begin{figure}[thbp]
\begin{center}
\includegraphics[width=8.5cm,clip,angle=0]{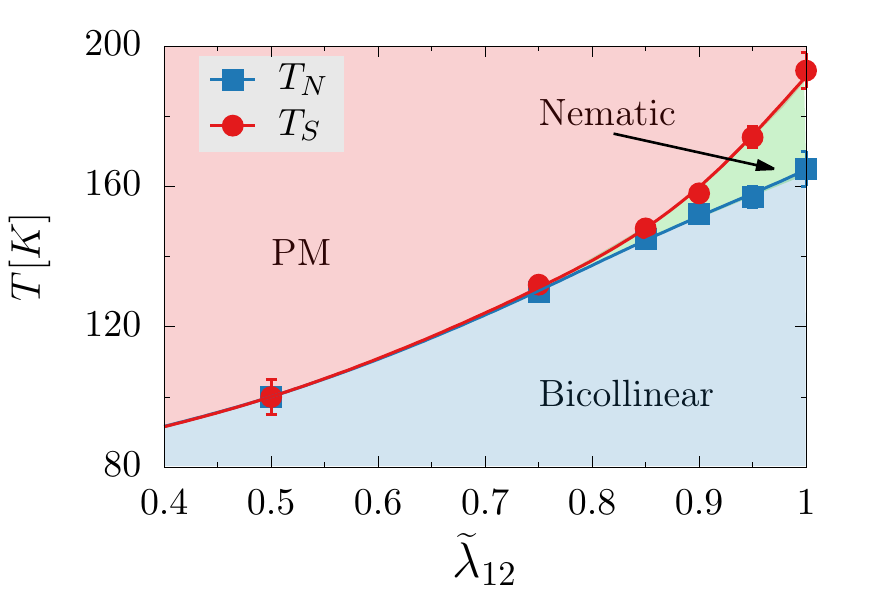}
\vskip -0.3cm
\caption{(color online) Phase diagram varying temperature and $\tilde\lambda_{12}$, 
for $\tilde g_{12}=0.24$, $J_{\rm H}=0.1$~eV, and $J_{\rm NN}$=$J_{\rm NNN}$=0.0.
Note the narrow temperature width of stability of the bicollinear-nematic state, similarly
as it occurs for the more standard $(\pi,0)-(0,\pi)$ nematic state~\cite{shuhua13}. For
values of $\tilde\lambda_{12}$ smaller than 0.75, our numerical accuracy does not allow us
to distinguish between $T_{N}$ and $T_{S}$.}
\label{pdiag}
\end{center}
\end{figure}

The phase diagram obtained as a function of the orbital-lattice coupling $\tilde\lambda_{12}$ and 
temperature is presented in Fig.~\ref{pdiag}. It can be seen that 
the region with $B_{2g}$ nematicity can be stabilized 
at robust  values of the orbital-lattice coupling. While a very narrow nematic
phase may exist at smaller values of this coupling, 
numerically we have been able to resolve the separation between the two critical temperatures only for 
$\tilde\lambda_{12}\ge 0.75$. 
As described in the previous sections, the separation between $T_{N}$ and $T_{S}$ monotonically 
increases with $\tilde\lambda_{12}$. 

\section{Discussion and Possible Physical Realizations}\label{disc} 

Our results have illustrated the 
possible existence of a nematic phase involving bicollinear short-range order, using as
explicit example a computational study of the
spin-fermion model incorporating 
the lattice distortions corresponding to  the iron telluride family. Previous investigations~\cite{bistripes} 
showed that the addition to the electronic 
spin-fermion model for pnictides of a coupling between a spin-nematic order parameter
with $B_{2g}$ symmetry and the monoclinic distortions of the  iron lattice
does induce the low-temperature monoclinic and spin bicollinear
state experimentally observed in FeTe. That result was 
remarkable because the spin-fermion model 
contains a tight-binding term that favors the $(\pi,0)$ and 
$(0,\pi)$ collinear states that 
arise from the nesting of the Fermi surface in weak coupling.
However, the $~\tilde g_{12}$ spin-lattice interaction, when sufficiently strong, 
can overcome these tendencies and stabilize the monoclinic bicollinear state. 

Here, we have included an additional 
orbital-lattice term with coupling strength $~\tilde\lambda_{12}$, 
involving the monoclinic lattice strain coupled 
to an orbital order parameter with $B_{2g}$ symmetry. By this procedure we have
shown that a novel nematic phase characterized 
by the breakdown of the lattice rotational symmetry between the two possible 
diagonal directions of the spin bicollinear state can be induced. 
In this new nematic phase, short-range spin- and orbital-nematic order develop accompanied 
by a lattice monoclinic distortion.

The model Hamiltonian studied here only allows 
us to show explicitly, as a matter of principle, 
that indeed the bicollinear-nematic state described above does occur
in computational studies once all of the many degrees of freedom and couplings 
are properly incorporated. But it is difficult to predict on what
specific material this subtle state will be realized in practice, thus we can only 
discuss some scenarios qualitatively. 
The possible splitting of $T_{N}$ and $T_{S}$ by electron doping 
was raised in~\cite{fernandes1}. However, spin-fermion model studies including doping but not quenched 
disorder  (i.e. in the ``clean'' limit) did not detect such a split, at least in
the doping range studied (Fig.~2 of ~\cite{chris15}). 
Another generic qualitative observation is that in the pnictides nematicity is observed
for the 1111 compounds even in the undoped limit~\cite{clarina}. Thus, to find the
$B_{2g}$ nematic phase discussed here
it may be necessary to synthesize materials with intercalated FeTe planes.

However, in our opinion
the most likely scenario to stabilize the proposed 
bicollinear-nematic regime in variations of the FeTe compound 
is by the chemical replacement of iron by other transition metal elements, 
thus simultaneously modifying
the electronic density as well as the amount of quenched disorder. 
In pnictides, replacing
Fe by Co, Ni, or Cu indeed leads to a wide nematic region. Our previous computational 
investigations using the spin-fermion model with doping and disorder~\cite{chris15} 
clearly showed that indeed by this procedure a $(\pi,0)$ nematic temperature range can be induced even in cases 
where $T_{N}$ and $T_{S}$ coincide in a first-order transition for the undoped parent compound,
as in the 122 family. Disorder plays a more important role than doping in this split~\cite{chris15}, as observed
experimentally as well~\cite{nni}.
To our knowledge the experimental investigations of (Fe,$X$)Te, with $X$ another
transition metal element, are very limited. We are aware of three main lines 
of investigations and conclusions:

{\it (i)} Copper doping of FeTe was studied in~\cite{FTCu,FTCu2} 
for two Cu concentrations using single crystals. For the case
Fe$_{1.06}$Cu$_{0.04}$Te the presence of strain was detected at 41$~$K upon cooling~\cite{FTCu}.
At lower temperatures approximately 36$~$K nearly-commensurate long-range bicollinear 
magnetic order occurs. The presence of two transitions seems in agreement with our prediction
of bicollinear nematicity. However, in~\cite{FTCu} it was argued that between 36$~$K and 41$~$K
the lattice distortion could be orthorhombic as in pnictides. The possible competition with
orthorhombic tendencies was theoretically addressed and reported in~\cite{bistripes}. 
This competition adds 
an extra  complication to the detection of the here predicted bicollinear-nematic state. 
 For the case FeCu$_{0.1}$Te only cluster glass behavior was found below 22~K, 
presumably due to disorder~\cite{FTCu}. 
Note that this glassy state could be nematic.

{\it (ii)} The case of Ni doping was reported for the compounds 
Fe$_{1.1-x}$Ni$_x$Te with $x=0, 0.02, 0.04, 0.08$, and $0.12$~\cite{FTNi}. Magnetization studies
show that $T_{N}$ decreases with increasing $x$ up to 0.04, while for $x=0.08,0.12$ a possible
spin glass transition was reported. In fact, neutron diffraction at $x=0.12$ found neither structural nor
magnetic transitions at low temperatures. Since this study focused on long-range magnetic order, the presence
of bicollinear nematicity is still possible.

{\it (iii)} Cobalt doping has also been recently studied via single crystals of 
Fe$_{1+y-x}$Co$_x$Te with $x=0,0.01,0.04,0.07,0.09,$ and $0.11$~\cite{FTCo}. 
In the range up to $x=0.07$ the antiferromagnetic transition systematically decreases.
For $x=0.09$ and larger the long-range order transition disappears. 

As a partial summary, the available experimental literature on (Fe,$X$)Te 
does not conclusively show
neither the presence nor absence of bicollinear-nematicity, and more work 
is needed to clarify this matter
now in the light of our present study. For example,
in the context of pnictides
the pioneering studies of Ba(Fe$_{1-x}$Co$_x$)$_2$As$_2$~\cite{fisher2010} reported
the resistivities vs. temperature 
along the $a$ and $b$ axes, highlighting their different behavior and substantial
differences particularly below $x=0.07$. Similar careful studies in the (Fe,$X$)Te 
context must be performed but focusing on the temperature evolution of the
resistivities along and perpendicular to the main spin diagonals in the bicollinear state,
as already performed for FeTe~\cite{reverse1,reverse2}.
In addition, recent 
inelastic neutron scattering studies of
nematicity in BaFe$_{1.935}$Ni$_{0.065}$As$_2$~\cite{dai-Ni} focused on the temperature
dependence of the intensity of the peaks at $(\pi,0)$ and $(0,\pi)$, reporting
their split at $T_{S}$ with cooling, followed by a collapse to zero
of the $(0,\pi)$ intensity at $T_{N}$. Similar
studies for $X$-doped FeTe ($X$=Cu,Ni,Co) should be carried 
for the temperature dependence of the 
neutron intensities corresponding to the $(\pi/2,\pi/2)$ and
$(\pi/2,-\pi/2)$ wavevectors.

We also would like to point out that our work confirms that magnetoelastic 
effects tend to estabilize the bicollinear state while in the absence of this kind
of coupling $Q$ plaquette or orthogonal double stripe order could be 
stabilized, which may be the case in FeTe with excess iron~\cite{ducatman,ducatman2}. 
In addition, in a recent publication~\cite{gzhang} a double-stage 
nematic bond-ordering above the bicollinear state was proposed, 
but this effect would be difficult to study numerically
due to the narrow range of the nematic phase.

\section{Conclusions}\label{conclu}

In this publication, based on simple symmetry observations and a concrete model Hamiltonian
numerical simulation, we have argued that the exotic bicollinear state known to be stable 
in FeTe admits a possible nematic state above the antiferromagnetic critical temperature. 
In other words, as discussed in the previous section, via chemical substitution  it is conceivable
that a split of the first-order transition of FeTe 
could be generated. Upon cooling, this would induce first 
a $T_{S}$, where the $B_{2g}$ monoclinic distortion is stabilized
and short-range spin and orbital order develops breaking 
the lattice rotational invariance, 
and second a $T_{N}$ at a lower temperature, where long-range bicollinear order is fully stabilized.
Experimentally finding this new exotic state not only would confirm the theoretical
prediction outlined here, but it would allow us to investigate to what extend nematic fluctuations are
needed to induce superconductivity.

\vspace{30pt}
\section*{Acknowledgments}
C.B. was supported by the National Science Foundation, under
Grant No. DMR-1404375. E.D., J.H., and A.M. were supported by the US Department of Energy, 
Office of Basic Energy Sciences, Materials Sciences and Engineering
Division.



\begin{references}
\bibitem{peaks} Notice that the magnetic structure 
factor for the state in Fig.~1a [1b] has peaks at both momentum $(\pi/2,-\pi/2)$  and
$(-\pi/2,\pi/2)$ [$(\pi/2,\pi/2)$ and $(-\pi/2,-\pi/2)$].
\bibitem{foot} The magnitude of the localized spins is set to $S_{\bf i}=1$ 
since its actual value can be absorbed into the Hamiltonian parameters.
\bibitem{intersite} Some calculations (not shown) for various sets of Heisenberg couplings, keeping
$J_{\rm NNN}/J_{\rm NN}=2/3$ constant, were performed for selected values of $~\tilde\lambda_{12}$ and $~\tilde g_{12}$. 
Only quantitative changes in the presented results were observed, while qualitatively they were similar. Thus,
the inclusion of Heisenberg couplings, which favor collinear order, does not destabilize the fragile bicollinear
nematic state, at least in the regime of robust lattice-orbital coupling. Details will be presented in future
publications.
\end{references}
\end{document}